\documentclass[preprint,12pt]{elsarticle}

\usepackage{natbib}
\usepackage{graphics,epsfig}
\usepackage{graphicx,epstopdf}
\usepackage{amssymb,amstext,amsmath}

\journal{European Journal of Mechanics A}

\begin{document}

\begin{frontmatter}

\title{Self-energy of dislocations and dislocation pileups}

\author{K. C. Le}

\address{Lehrstuhl f\"{u}r Mechanik - Materialtheorie, Ruhr-Universit\"{a}t Bochum, \\ D-44780 Bochum, Germany}

\begin{abstract} A continuum model of dislocation pileups that takes the self-energy of dislocations into account is proposed. An analytical solution describing the distribution of dislocations in equilibrium is found from the energy minimization. Based on this solution we show (i) the existence of a critical threshold stress for the equilibrium of dislocations within a double pileup, and (ii) the existence of a non-linear regime in which the number of dislocations in a double pileup does not scale linearly with the resolved external shear stress, contrary to the classical double pileup model.
\end{abstract}


\end{frontmatter}

\section{Introduction}

In recent years there is a substantial amount of literature dealing with the dislocation pileups in crystals within the continuum approach (see, e.g., \citep{Berdichevsky-Le07,Gurtin2002,Gurtin2007,Kaluza2011torsion,Kochmann08a,Kochmann08b,Le2012polygonization,Le2013on,Le08a,Le08b,Le2009plane,Ohno2007} and the references therein). The proposed models turn out to be quite useful as they can predict the size effect for the yield stress, which agrees quite well with the experimental data \citep{Ohno2007} but however does not confirm the well-known empirical law formulated by \citet{Hall1951} and \citet{Petch1953}. A traditional explanation of the Hall-Petch relation, based on the classical dislocation pileup model considered in \citep{Leibfried51}, is that dislocation pile-ups serve to enhance the stress felt at grain boundaries. However, the Leibried's dislocation pileup model differs from the contemporary continuum models \citep{Ohno2007,Berdichevsky-Le07,Kaluza2011torsion,Kochmann08a,Kochmann08b,Le2012polygonization,Le2013on,Le08a,Le08b,Le2009plane} in two important aspects: firstly the absence of a threshold stress for dislocation nucleation, and secondly the absence of a finite-sized dislocation-free region. This is the motivation for us to reconsider the Leibfried's model in order to resolve this discrepancies. 

Leibfried's one-dimensional theory of dislocation pileups is based on the force equilibrium: the Peach-Koehler resultant force acting on a dislocation produced by applied external stress and by other dislocations must vanish \citep{Eshelby51a,Leibfried51}. Within the continuum approximation one can deduce from here the well-known integral equation \citep{Leibfried51}, provided the dislocation density is not zero. This equation need not be satisfied in a dislocation-free zone. In his now classical paper \citet{Leibfried51} mentioned that the natural way to find the stable equilibrium distribution of dislocations if such a zone occurs is to use the variational principle of minimum energy: among all admissible distributions of dislocations the true stable distribution minimizes energy of the crystal (cf. with the LEDS-hypothesis formulated in \citep{Hansen1986}). To the best of the author's knowledge, the one-dimensional continuum model of dislocation pileups based on the energy minimization taking the self-energy of dislocations into account has not yet been proposed. The aim of this short paper is to fill this gap. We start from the expression for the energy of crystal containing an array of dislocations which includes also the self-energy of dislocations. The self-energy density is proportional to the absolute value of the dislocation density, so this does not change the resultant force acting on the dislocation as well as the integral equation except the forces acting at the tails of the pileups. However, this self-energy influences the stable equilibrium distribution of dislocations essentially. We will show that the density of dislocations is identically zero if the applied stress is found below some critical value called the yield (or threshold) stress. Besides, the number of dislocations depends on the applied stress non-linearly, in contrast to the classical theory. The results of the proposed theory agree quite well with those of the continuum models of dislocation pileups obtained in \citep{Berdichevsky-Le07,Kaluza2011torsion,Kochmann08a,Kochmann08b,Le08a,Le08b,Le2009plane}. A continuum model of dislocation pile-ups taking into account the Frank-Read source proposed in \citep{Friedman98} (see also \citep{Chakravarthy2010}) leads to the similar results although it does not have the energetic structure.

In the next Section we present the continuum model of dislocation pileup accounting for the self-energy of dislocations. In Section 3 the energy minimization problem is solved and the related property of the solution is discussed. Finally, Section 4 concludes the paper.

\section{Continuum model of dislocation pileup}

\begin{figure}[htb]
	\centering
	\includegraphics[width=6cm]{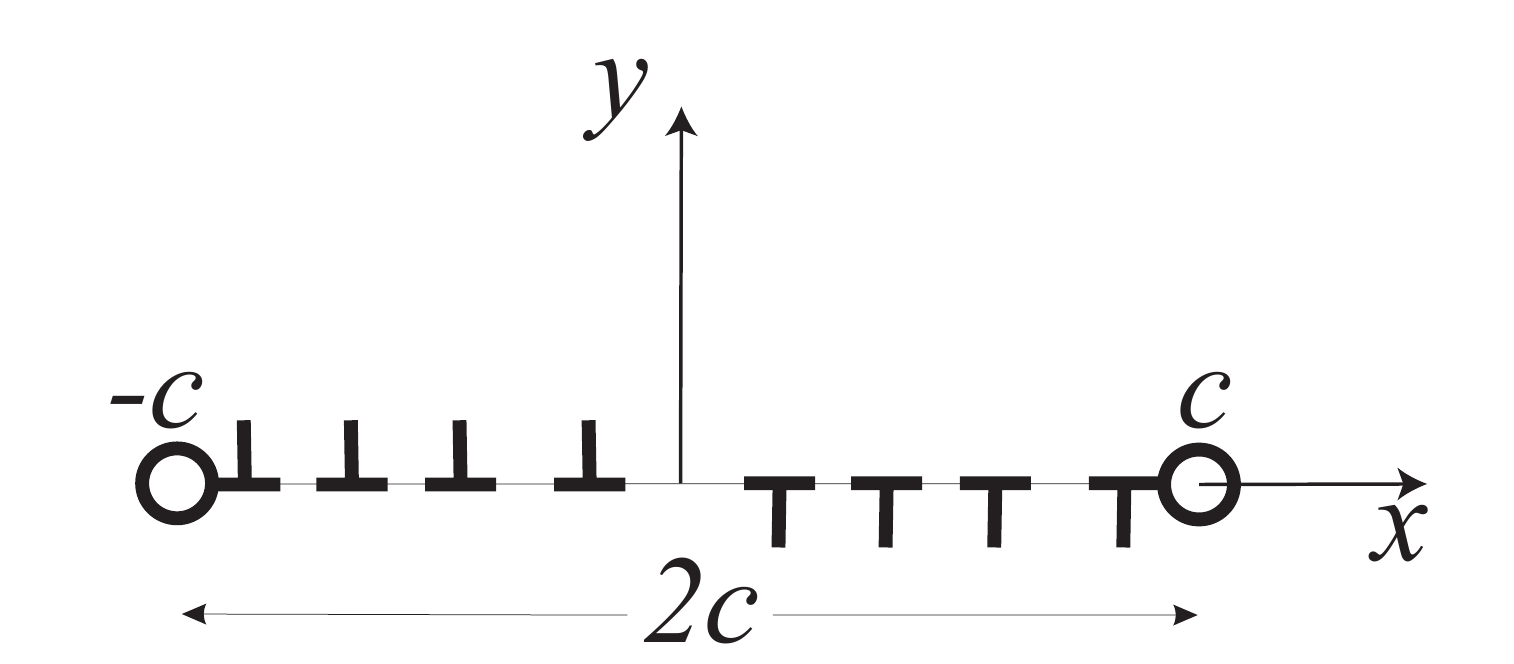}
	\caption{Double pileup of edge dislocations}
	\label{fig:a}
\end{figure}

\begin{figure}[htb]
\centering \epsfig{file=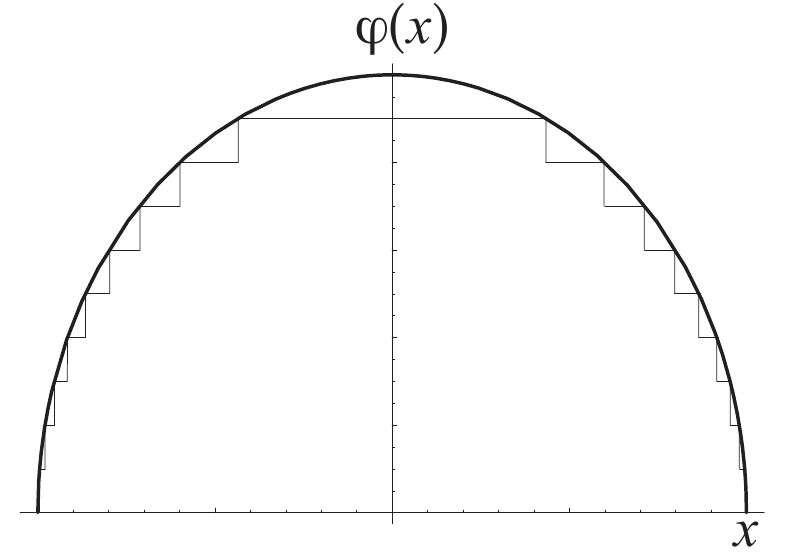,width=5.5cm} \caption{Continuum approximation of the closure failure} \label{fig:pileup3}
\end{figure}

Consider the plane strain problem of an infinite crystal which is uniformly loaded by a shear stress $\tau $ applied at infinity. Under this loading condition a linear array of equal number of positive and negative edge dislocations may occur on the slip line which is chosen to be the $x$-axis (see Fig.~\ref{fig:a}). The dislocation lines are parallel to the $z$-axis, while their Burgers' vectors are directed along the $x$-axis. We assume that there are two obstacles (like two inclusions or grain boundaries) at $x=\pm c$ so that dislocations are confined to stay in the interval $L=(-c,c)$ of the $x$-axis. In the continuum limit we may replace the sum of many closure failures induced by dislocations in form of step functions by a smooth function $\varphi (x)$ (see Fig.~\ref{fig:pileup3}). The obstacles impose the following constraints on this function 
\begin{equation}\label{1.3}
\varphi (\pm c)=0.
\end{equation} 
Now we present the resultant inverse plastic distortion \citep{Le2010} in the form
\begin{equation*}
-\beta _{xy}=\varphi (x)\delta (y),
\end{equation*}
with $\delta (y)$ being the Dirac-delta function. Differentiating this equation with respect to $x$ we obtain 
\begin{equation}\label{pileup3}
-\beta _{xy,x}=\varphi ^\prime (x)\delta (y).
\end{equation}
The interpretation of \eqref{pileup3} is quite simple: if we integrate this equation over a circle $C$ with the middle point at $x=-c$ and the radius $c+x$, then
\begin{equation*}
-\int_{C}\beta _{xy,x}\, dxdy=-\int_{\partial C}\beta _{xy}dy=\int_{-c}^{x} \varphi ^\prime (\xi )\, d\xi =\varphi (x).
\end{equation*}
Thus, we get the closure failure of an amount $\varphi (x)$ which should be equal to the net Burgers' vector of all dislocations within the interval $(-c,x)$. Denoting the dislocation density by $\rho (x)$, we get
\begin{equation}\label{pileup}
\varphi (x)=b\int_{-c}^{x}\rho (\xi )\, d\xi \quad \Rightarrow \varphi ^\prime (x)=b\rho (x).	
\end{equation}
So, it is extremely useful to think of this double pileup of dislocations as a mode II crack\footnote{In fact, various crack problems have been solved within this continuum model (see, for example, \citep{Lardner1974,Weertman1996}).} and to interpret the closure failure $\varphi (x)$ as the crack opening which is shown schematically in Fig.~\ref{fig:pileup3}. 

The energy of this system can be found directly from the energy of crystal containing continuously distributed dislocations in the interval $(-c,c)$ of the $x$-axis. Indeed, for the plane strain state the energy functional per unit depth of the crystal reads
\begin{equation*}
I[\mathbf{u}(x,y)]=\int_{A}\phi (\varepsilon	_{\alpha \beta }-\varepsilon ^p_{\alpha \beta })\, dxdy-\int_{\partial A}\tau _\alpha u_\alpha \, ds,
\end{equation*}
where $A$ is the cross-section area of the crystal in form of a cylinder, $\phi (\varepsilon ^e_{\alpha \beta })$ the free energy density, $\varepsilon ^p_{\alpha \beta }$ the symmetric part of the plastic distortion, while $\tau _\alpha $ is the external traction acting at the boundary  $\partial A$. Substituting $\phi =\frac{1}{2}\sigma _{\alpha \beta }(u_{\alpha ,\beta }-\beta _{\alpha \beta })$ into this energy functional and integrating the term $\sigma _{\alpha \beta }u_{\alpha ,\beta }$ by parts using the equilibrium equation and the boundary conditions, we can show that this term is canceled out with the last term, so
\begin{equation*}
I=-\frac{1}{2}\int_{A}	\sigma _{xy}\beta _{xy}\, dxdy=\frac{1}{2}\int_L \sigma _{xy}(x)\varphi (x)\, dx.
\end{equation*}
Let us first exclude the self-stress (causing the self-energy of dislocations) and neglect the influence of the boundary of crystal by assuming that $A$ occupies the whole $(x,y)$-plane. Then we can present the shear stress $\sigma _{xy}(x)$ on the $x$-axis in the form
\begin{equation}\label{pileup5}
\sigma _{xy}(x)= D {\bf -}\!\!\!\!\!\!\int_L
\frac{\varphi ^\prime (\xi )}{\xi -x}\, d\xi +\tau , \qquad  D = \frac{\mu
}{2\pi (1-\nu )}.
\end{equation}
The first term in this formula gives the stress field induced by the continuous distribution of dislocations excluding those in the neighborhood of $x$ (the integral ${\bf -}\!\!\!\!\!\int^{}_{}$ in \eqref{pileup5} is defined as Cauchy's principal value), the second term corresponds to the stress field caused by the external shear stress $\tau $. Substituting \eqref{pileup5} into the energy functional and integrating the first term by parts, we obtain finally
\begin{equation}\label{pileup10}
I[\varphi (x)]=-\frac{D}{2}\int_L\int_L \ln |x-\xi |\varphi ^\prime (x)\varphi ^\prime (\xi )\,d\xi \, dx -\int_L \tau \varphi (x)\, dx.
\end{equation}
The double integral in \eqref{pileup10} is the symmetric form of the quadratic functional which should be understood in the following sense
\begin{equation}\label{1.2}
-\int_{L} \int_{L} \ln |x-\xi |\, \varphi ^\prime (x)\varphi
^\prime (\xi)\, dx d\xi =-\int_{L} {\bf -}\!\!\!\!\!\!\int_L \frac{\varphi
^\prime (\xi) }{\xi -x}\,d\xi \varphi (x)\, dx .
\end{equation}
It is interesting to note that this quadratic functional is positive definite and obeys an inequality similar to that of Wirtinger (see \citep{Le06a}). Note also that, since the left-hand side is invariant with respect to the scaling $x\to x/c$ due to the boundary condition \eqref{1.3}, the arguments in the log-kernel can be regarded as dimensionless.

Functional \eqref{pileup10} is the energy functional of the classical continuum dislocation theory in which, as we have seen from the previous derivation, the self-stress of dislocations is excluded. This means that the self-energy of dislocations is neglected. Let us now take the self-energy of dislocations into account by the following deliberations. It is well-known that the self-energy of one dislocation (per unit length) does not depend on its sign and equals $\kappa \mu b^2$, where $\kappa =\ln (4R/b)/(4\pi (1-\nu ))$, with $R$ being the distance of dislocation to the boundary of crystal and $\nu $ the Poisson ratio \citep{Hirth1968}. Since the dislocation density is $\rho (x)=\frac{1}{b}\varphi ^\prime (x)$ and since the self-energy does not depend on the sign of dislocations, the total self-energy of dislocations is given by
\begin{equation*}
\kappa \mu b\int_{L} |\varphi^\prime (x)| dx.
\end{equation*}
Adding this energy to the functional \eqref{pileup10} we get finally
\begin{equation}\label{1.1}
I[\varphi (x)]=-\frac{D}{2} \int_{L} \int_{L} \ln |x-\xi |\,
\varphi ^\prime (x) \varphi ^\prime (\xi) dx d\xi -\int_{L}\tau
\varphi (x) dx+\kappa \mu b\int_{L} |\varphi^\prime (x)| dx,
\end{equation}
We formulate the following variational principle: the stable equilibrium distribution of dislocations corresponds to the minimizer $\varphi (x)$ of functional \eqref{1.1} under the constraints \eqref{1.3}. The existence of minimizer is guaranteed by the above mentioned inequality for the quadratic functional \eqref{1.2} provided in \citep{Le06a}. 

\section{Energy minimization}
Symmetry of the problem implies that $\varphi (x)$ is even and $\varphi ^\prime (x)$ is odd. It can be shown that the presence of $|\varphi ^\prime (x)|$ in the last term of the energy functional \eqref{1.1} causes the minimizer to have the dislocation-free zone as observed in \cite{Berdichevsky-Le07}. Therefore we assume that there is a length $a<c$ such that
\begin{equation}\label{2.0}
\varphi ^\prime (x)=
  \begin{cases}
    >0 & \text{for $x\in L_-=(-c,-a)$}, \\
    <0 & \text{for $x\in L_+=(a,c)$}, \\
    0 & \text{otherwise}.
  \end{cases}
\end{equation}
Thus, there are two boundary layers where positive and negative dislocations pile up against the obstacles and the dislocation-free zone in-between. Let us first fix the constant value of $\varphi (x)$ in $(-a,a)$, $\varphi (a)$, as well as the length $a$. Varying functional \eqref{1.1} and taking into account \eqref{1.3} we get for the first variation
\begin{equation*}
\delta I=\int_{L} \left[ -D{\bf -}\!\!\!\!\!\!\int_{L_-}
\frac{\varphi ^\prime (\xi)}{\xi-x}\, d\xi -D{\bf
-}\!\!\!\!\!\!\int_{L_+} \frac{\varphi ^\prime (\xi)}{\xi-x}\,
d\xi -\tau \right] \delta \varphi (x)\, dx.
\end{equation*}
There is no contribution of the last integral in \eqref{1.1} due to the fact that it is equal to $2\kappa \mu b \varphi (a)$ which is fix. Since $\delta \varphi (x)$ may be chosen arbitrarily for $x\in L_\pm $, $\varphi (x)$ should satisfy the singular integral equation
\begin{equation}\label{2.1}
{\bf -}\!\!\!\!\!\!\int_{L_-} \frac{\varphi ^\prime (\xi)} {\xi
-x}\, d\xi +{\bf -}\!\!\!\!\!\!\int_{L_+} \frac{\varphi ^\prime
(\xi)} {\xi -x}\, d\xi =-\frac{\tau }{D} \quad \text{for $x\in
L_\pm $}.
\end{equation}
Now, it is not difficult to show that the variation of $a$ leads to the boundary conditions
\begin{equation}\label{2.2}
\varphi ^\prime (\pm a)=0,
\end{equation}
which guarantee the continuity of the dislocation density. Thus, equation \eqref{2.1} is subjected to the boundary conditions \eqref{1.3} and \eqref{2.2}.
The solution of \eqref{2.1}, \eqref{1.3}, and \eqref{2.2}, found in \citep{Leibfried51} (see also \citep{Voskoboinikov2007,Hall2010}), reads
\begin{equation*}
\varphi ^\prime (x)=
  \begin{cases}
    \mp \frac{\tau }{\pi D}\sqrt{\frac{x^2-a^2}{c^2-x^2}} & \text{for $x\in
L_\pm $}, \\
    0 & \text{otherwise}.
  \end{cases}
\end{equation*}
From here one easily finds $\varphi (x)$ in terms of the elliptic integrals (see \citep{Gradshteyn00}, p. 276)
\begin{equation}\label{2.4}
\varphi (x)=  \begin{cases}
    \frac{\tau }{\pi D}[cE(\lambda ,q)-\frac{a^2}{c}F(\lambda ,q)]
    & \text{for $x\in L_\pm $}, \\
    \frac{\tau }{\pi D}[cE(q)-\frac{a^2}{c}K(q)] &
    \text{otherwise},
    \end{cases}
\end{equation}
where
\begin{equation*}
  \lambda = \arcsin \sqrt{\frac{c^2-x^2}{c^2-a^2}},\quad
  q=\frac{\sqrt{c^2-a^2}}{c}.
\end{equation*}
In formula \eqref{2.4} $F(\lambda ,q)$ and $E(\lambda ,q)$ are the incomplete elliptic integrals of the first and second kind, respectively, while $K(q)$ and $E(q)$ are the complete elliptic integrals of the first and second kind. Mention that $\varphi (a)/b$ gives the total number of dislocations of either sign in the pileups (cf. \eqref{pileup}).

Up to now $\varphi (a)$ is an unknown quantity. It can be found by minimizing energy as function of $a$. Substituting the solution \eqref{2.4} into the energy functional \eqref{1.1} and using \eqref{2.1} we obtain
\begin{equation}\label{2.5}
I(\tau ,a)=-\frac{\tau }{2}\int_{L}\varphi (x)\, dx+2\mu \kappa
b\varphi (a)=-\tau \int_{L_+}\varphi (x)\, dx-\tau \varphi
(a)a+2\mu \kappa b\varphi (a).
\end{equation}
The direct consequence of \eqref{2.5} is that for $\tau <2\mu \kappa b/c$ the minimum of energy is achieved at the end-point $a=c$ giving $I=0$. Indeed, for $a<c$ we have $\varphi (x)<\varphi (a)$, so, $I(\tau ,a)>0$ in this case and the minimum is achieved at $a=c$. Thus, the threshold stress $\tau _c=2\mu \kappa b/c$. Note that this threshold stress is inversely proportional to the size of the grain exhibiting clearly the size effect (cf. with \citep{Ohno2007}).

\begin{figure}[htb]
	\centering
	\includegraphics[width=8cm]{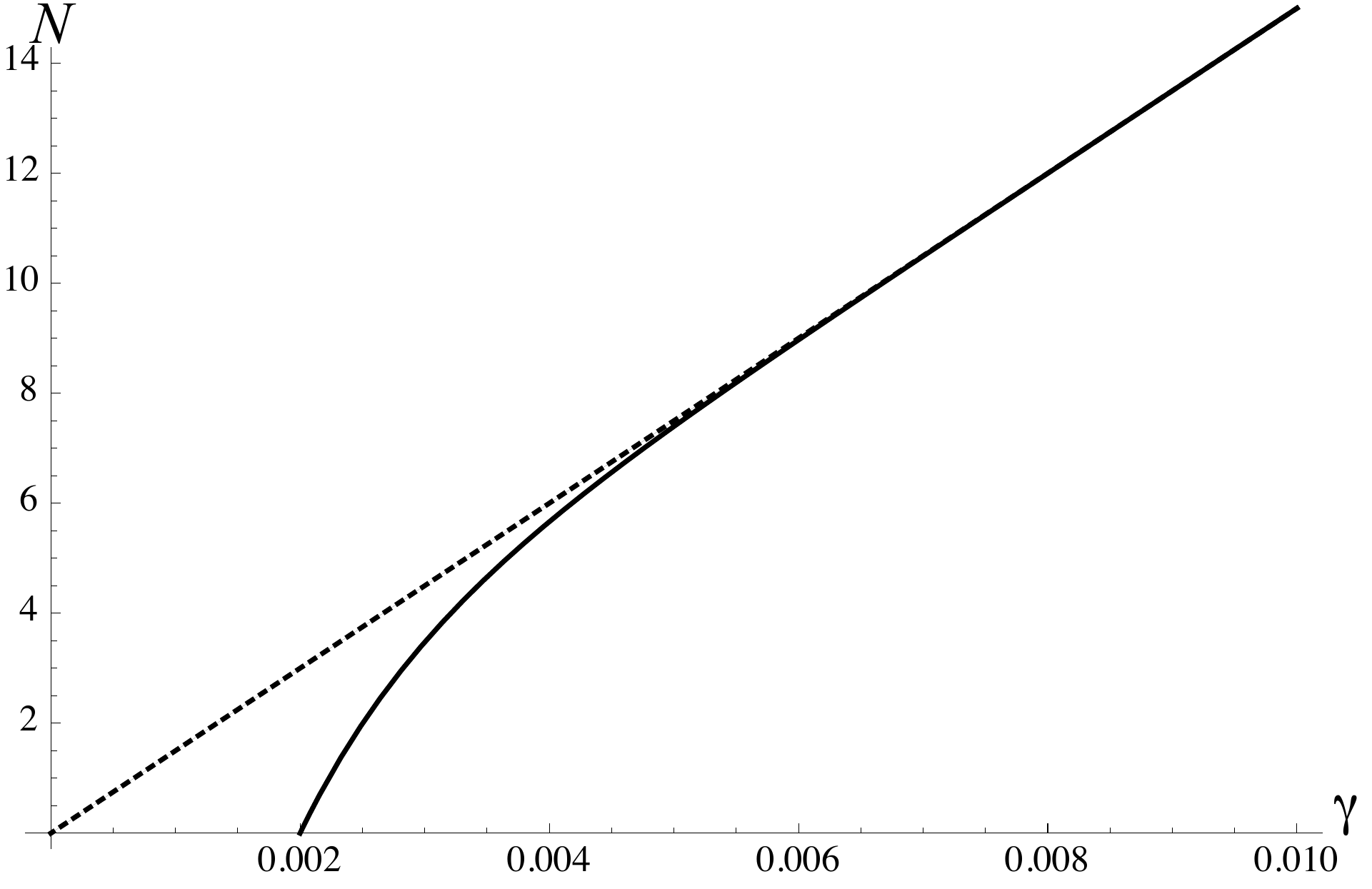}
	\caption{Graph of function $N(\gamma )$: a) bold line: theory with self-energy of dislocations, b)
dashed line: classical theory}
	\label{fig:1}
\end{figure}

In general the integral in \eqref{2.5} cannot be found in closed analytical form. However it is easy to find minimum of \eqref{2.5} with respect to $a$ numerically. For this purpose let us introduce the following dimensionless variable and quantities
\begin{equation*}
\zeta =\frac{x}{c},\quad \Phi (\zeta )=\varphi (\zeta c)/c,\quad
\alpha =a/c,\quad \beta =b/c,\quad \gamma =\tau /\mu ,\quad
\bar{I}=\frac{I}{\mu c^2}.
\end{equation*}
Then function $\Phi (\zeta )$ is given by
\begin{equation*}
\Phi (\zeta )=  2\gamma (1-\nu )[E(\lambda ,q)-\alpha ^2F(\lambda
,q)] \quad \text{for $\zeta \in (\alpha ,1)$},
\end{equation*}
where
\begin{equation*}
\lambda =\arcsin \sqrt{\frac{1-\zeta ^2}{1-\alpha ^2}},\quad
q=\sqrt{1-\alpha ^2}.
\end{equation*}
Formula \eqref{2.5} becomes
\begin{equation*}
\bar{I}(\gamma ,\alpha )=-\gamma \int_{\alpha }^1\Phi (\zeta)\,
d\zeta -\gamma \Phi (\alpha )\alpha +2\kappa \beta \Phi (\alpha ).
\end{equation*}
We evaluate this function numerically using {\it Mathematica}. The minimum of $\bar{I}(\gamma ,\alpha )$ with respect to $\alpha $ is also sought numerically. After finding $\alpha _*$ at which energy reaches the minimum, we find the total number of dislocations of either sign in the pileups by (cf. \eqref{pileup})
\begin{equation*}
N=\frac{\Phi (\alpha _*)}{\beta }=\frac{2\gamma (1-\nu )}{\beta }
[E(q(\alpha _*))-\alpha _*^2K(q(\alpha _*))].
\end{equation*}
In Fig.~\ref{fig:1} the graph of $N$ versus the shear strain $\gamma $ is plotted, where, for comparison, the dashed straight line $N=2\gamma (1-\nu )/\beta $ obtained by the classical theory is also shown. For the numerical calculation we took $\beta =0.001$, $\kappa =1.0$, $\nu =0.25$. It is seen that for $\gamma
<2\kappa \beta $ no dislocations can be formed. The curve $N(\gamma )$ obtained from the energy minimization is a non-linear function of $\gamma $, but for $\gamma >6\kappa \beta $ it approaches quickly the straight line $N=2\gamma (1-\nu )/\beta $.

\begin{figure}[htb]
	\centering
	\includegraphics[width=10cm]{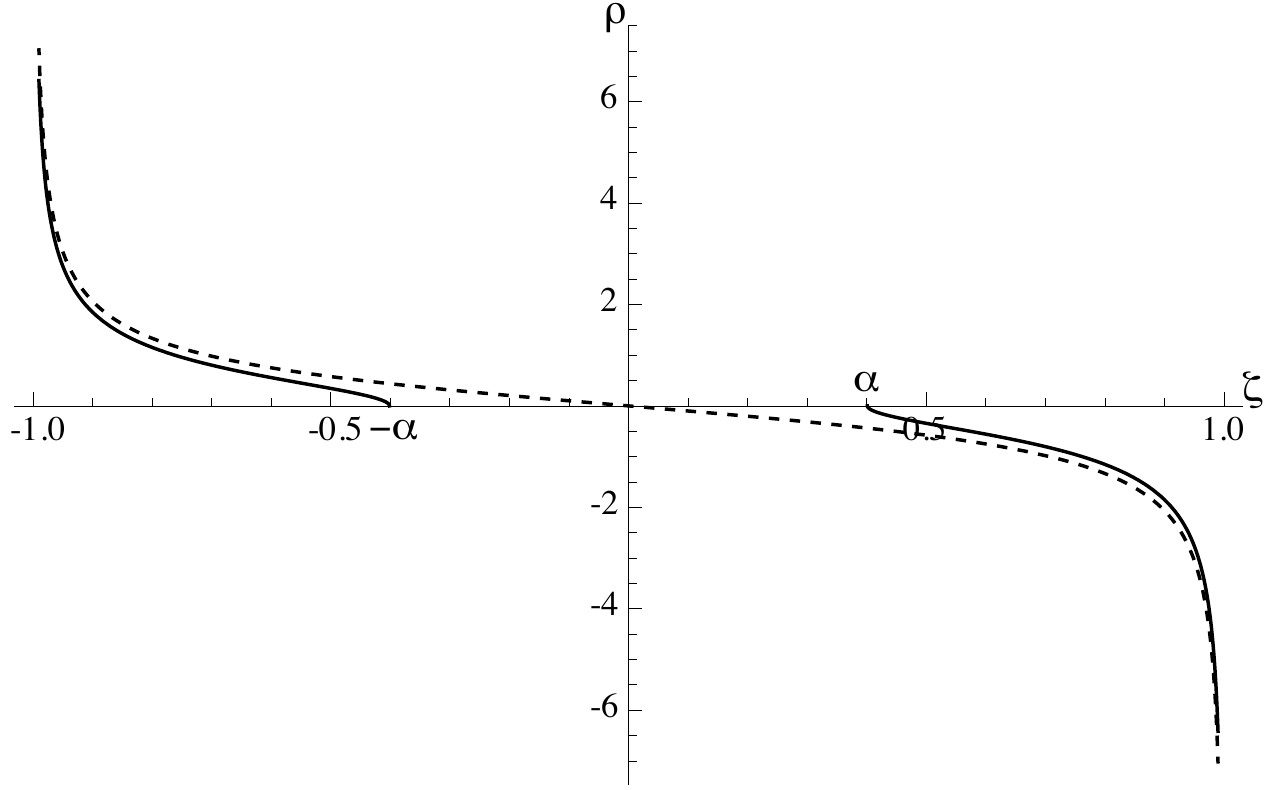}
	\caption{Normalized dislocation density $\rho (\zeta )$: a) bold line: theory with self-energy of dislocations, b) dashed line: classical theory}
	\label{fig:2}
\end{figure}

Fig.~\ref{fig:2} shows the normalized dislocation density $\rho =\pi Dc\varphi ^\prime /\tau $ as function of the dimensionless coordinate $\zeta =x/c$ for $\gamma =0.003$ which, in contrast to the classical distribution $-\zeta /\sqrt{1-\zeta ^2}$ (presented by the dashed line), depends on the stress (or the shear strain $\gamma $) through the half-length of the dislocation-free zone $\alpha $. However, this haft-length approaches quickly zero for $\gamma >6\kappa \beta $, so the distribution of dislocations approaches quickly that of the classical theory at large $\gamma $.

\section{Conclusion}
It is shown in this paper that the account of self-energy of dislocations leads to the existence of the threshold stress at which the dislocations nucleate. The total number of dislocations depends non-linearly on the applied shear stress, however it approaches quickly that of the classical theory for the stress three times larger than the threshold value.

\bigskip
\noindent {\it Acknowledgments}

The financial support by the German Science Foundation (DFG) through the research project LE 1216/4-2 is gratefully acknowledged.

\end{document}